\documentclass[prb,twocolumn,groupedaddress]{revtex4}
\usepackage{graphicx}
\begin{document}
\title{Preferential attachment during the evolution of a potential energy landscape} 
\author{Claire P. Massen}
\affiliation{University Chemical Laboratory, Lensfield Road, Cambridge CB2 1EW, United Kingdom}
\author{Jonathan P.~K.~Doye}
\affiliation{Physical and Theoretical Chemistry Laboratory, University of
  Oxford, South Parks Road, Oxford OX1 3QZ, United Kingdom}
\date{\today}

\begin{abstract}
It has previously been shown that the network of connected minima on a potential
energy landscape is scale-free, and that this reflects a power-law distribution
for the areas of the basins of attraction surrounding the minima.
Here, we set out to understand more about the physical origins of these puzzling
properties by examining how the potential energy landscape of a 13-atom
cluster evolves with the range of the potential.
In particular, on decreasing the range of the potential the number of stationary points increases and thus the landscape becomes rougher and 
the network gets larger.
Thus, we are able to follow the evolution  of the potential energy landscape 
from one with just a
single minimum to a complex landscape with many minima and a scale-free pattern
of connections.
We find that during this growth process, 
new edges in the network of connected minima preferentially attach to 
more highly-connected minima, thus leading to the scale-free character. 
Furthermore, minima that appear when the range of the potential is shorter 
and the network is larger have smaller basins of attraction.
As there are many of these smaller basins because the network grows 
exponentially, the observed growth process thus also gives rise to a power-law 
distribution for the hyperareas of the basins.
\end{abstract}

\maketitle

\section{Introduction}

A potential energy landscape (PEL) is a high-dimensional surface describing how the potential energy of a system varies with the coordinates of each atom in the system.\cite{Landscapes}
Since PELs control the structure, thermodynamics and dynamics of a system, there has been an ongoing research programme that has sought to understand how the behaviour of a system can be related back to features of a complex PEL.
This approach has led to new insights into protein folding \cite{Bryngelson95} and the nature of the glass transition. \cite{Debenedetti01}
In these applications the emphasis is on how PELs differ, for example between proteins that are good and bad folders or liquids that are fragile and strong.
In this paper by contrast, we are interested in more universal properties of high dimensional PELs.

\begin{figure}
\includegraphics[width=8.6cm]{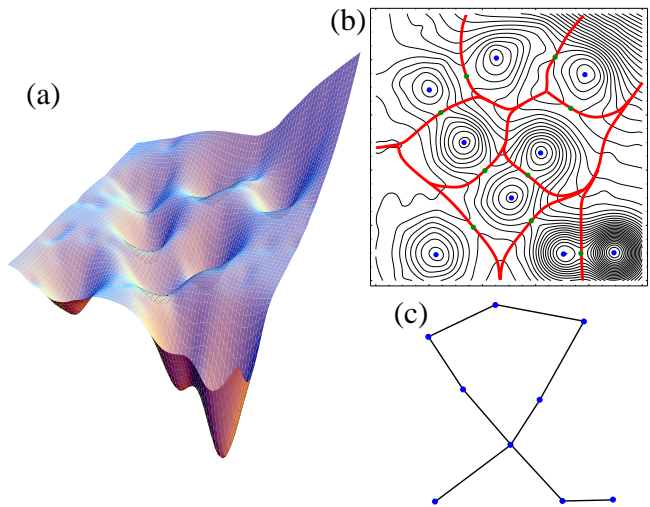}
\caption{(Colour online).
(a) A model two-dimensional energy surface.
(b) A contour plot of this surface
illustrating the inherent structure partition of the configuration space
into basins of attraction surrounding minima.
The basin boundaries are represented by thick lines,
and the minima and transition states by dots.
(c) The resulting representation of the landscape as a network.
}
\label{fig:pes}
\end{figure}

When characterizing PELs the focus is often on the stationary points of these surfaces.
Some fundamental properties associated with stationary points of PELs are well established.
For example, the number of stationary points increases exponentially with system size. \cite{Tsai93,Doye02b,Stillinger99}
The energy distribution of both minima and transition states follows a Gaussian distribution, as has been seen empirically, \cite{Buchner99,Sciortino99} and predicted theoretically from the central limit theorem. \cite{Heuer00}
More recently, network-derived insights have been obtained for PELs. \cite{Doye02,Doye05b,Massen05,Doye05c,Massen05c}
To obtain a network representation of a PEL, each minima is mapped to a node in the network and each transition state to an edge, as in Fig.~\ref{fig:pes}.
The resulting ``inherent structure'' network \cite{Doye02, Doye05b} provides
a dynamically motivated description of the connectivity and organization of 
the PEL, since at sufficiently low temperatures, a system vibrates within the basin of attraction of a minimum, or inherent structure,\cite{Stillinger84} before moving to another via a transition state valley.\cite{Schroder00} 

For small Lennard-Jones clusters, these networks have been shown to have 
both scale-free and small world properties.\cite{Doye02,Doye05b}
The characteristic of a scale-free network is a power-law tail to the 
degree distribution, \cite{Barabasi99} 
where the degree of a node is the number of edges connected to it,
and the characteristic of a small-world network is a logarithmic increase
in the average separation between nodes with network size.\cite{Watts98}
These two properties are related since the high-degree hubs present in
scale-free networks connect up the network, 
making the path lengths between nodes very short.\cite{Cohen02}
Recently, a wide variety of networks have been shown to be scale 
free,\cite{Newman03,Dorogovtsev03}
including the world wide web, \cite{Albert99, Faloutsos99} social networks \cite{Newman01c, Redner98} and biological networks. \cite{Jeong00, Jeong01}
Typically, the exponent of the power-law lies between $-2$ and $-3$, as is the
case for the inherent structure networks, which have an exponent of 
approximately $-2.78$.\cite{Doye02,Doye05b} 
Interestingly, networks based on the configuration space of a protein, \cite{Rao04,Gfeller07} although defined in a more coarse-grained way than
the inherent structure networks, are also scale-free, suggesting that this 
feature could be a universal property of PELs.

A popular explanation for the scale-free character of many of these networks
is in terms of preferential attachment during the growth of these networks.\cite{Barabasi99}
In preferential attachment, new nodes added to the network are more likely to 
attach to old nodes with high degree.
Therefore, the degree of high-degree nodes increases at a faster rate, a `rich get richer' growth process, and leads to a power-law degree distribution. 

By contrast, a PEL is static---it is just determined by the form of the potential and the system size.
Instead, for PELs, the scale-free nature seems to be related to the 
distribution of basin areas.  Each minimum on the PEL is surrounded by a basin 
of attraction, a region of configuration space where following the gradient downhill will lead to the same minimum.
Low-energy minima have large basins of attraction,\cite{Doye98,Massen05b} 
and so unsurprisingly they have high degree because they can fit many 
neighbours along their long boundaries.
A broad degree distribution thus reflects a broad basin area distribution. 

\begin{figure}
\centerline{\includegraphics[width=7.5cm]{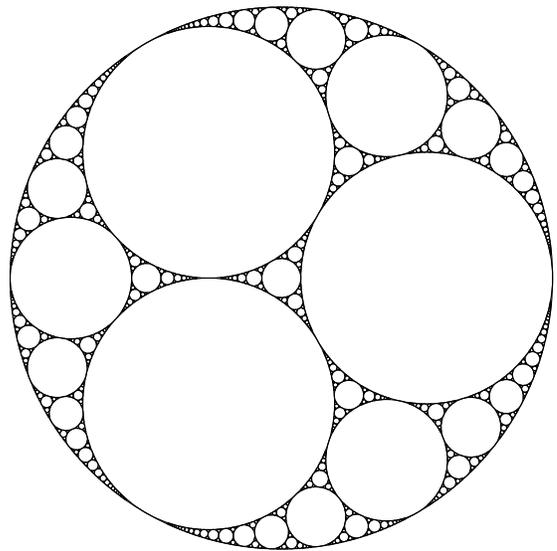}}
\caption{
A two-dimensional Apollonian packing.
Space is filled with different sized disks, starting from an initial configuration where the three larger disks are placed within the bounding circle.
The packing is generated iteratively, with the largest disk possible added to each gap at each iteration.
This process is continued ad infinitum, thus filling the space with successively smaller and smaller disks.
}
\label{fig:apollo}
\end{figure}

This connection between area and degree is more explicit in Apollonian networks, which are model spatial, scale-free networks \cite{Andrade05,Doye05} that have similar topological properties to inherent structure networks. \cite{Doye05,Doye05c}
These Apollonian networks are based on Apollonian packings in which all of space is tiled by hyperspheres (see Fig.~\ref{fig:apollo} for a two-dimensional example).
Each hypersphere gives rise to a node in the network, and two nodes are connected if the corresponding hyperspheres touch.
The packing is constructed iteratively and at each step in the generation of the packing each interstice is filled by the largest possible hypersphere that just touches the surrounding hyperspheres.
The fractal, self-similar character of the packings leads both to the scale-free topology of the Apollonian networks, and a power-law distribution for the hypervolumes of the hyperspheres.

The Apollonian packings and networks suggest how the basins of attraction on a PEL might have to tile configuration space in order to generate a scale-free inherent structure network.
Namely, the tiling would need to be hierarchical with larger basins surrounded by smaller basins, which are in turn surrounded by smaller basins, and so on.
Interestingly, when the basin area distribution has been probed for model liquids, a power-law distribution has been found with the exponent expected from analogy to the Apollonian packings, \cite{Massen05} thus suggesting that the inherent structure networks for these systems are scale-free (note that it is effectively impossible to obtain these networks directly due to the high 
dimensionality of their configuration space).
Although these results and ideas show how the topology of the inherent structure networks can be understood in terms of how the basins of attraction tile configuration space, why this tiling should be hierarchical, fractal and Apollonian-like is still a puzzle.

Recently, using a simple model we explored whether these features could 
simply be driven by the variation in the energies of the minima.\cite{Massen07}
In this `eggbox' model, we took the Gaussian distribution of energies of the 
minima as a starting point, and examined how the PEL landscape evolved (e.g.\ 
through the swallowing up of higher energy minima by deep basins) as
the variance of the Gaussian increased.
Although the resulting area distribution was considerably broader than for 
a flat landscape where all the minima have the same energy, these changes
were not sufficient to genearate a power law tail for the degree or basin area
distributions.

Here, we provide an alternative perspective on why the inherent structure networks for PELs are scale-free, and, as we will see, hence obtain insights into the nature of the division of configuration space into basins of attraction.
Although, as already mentioned, a PEL is static, there may be ways to envisage 
the generation of the network as a quasi-growth process, and
hence to apply the ideas of preferential attachment.
Recently, we looked at the role played by the energetic ordering of minima. \cite{Massen05c}
Here, we wish to look at how the PEL (and the inherent structure network) evolves on going from a simple landscape with one minimum to a complex one with many ($N$) minima and a scale-free topology.
The reverse of this process, smoothing a complex landscape so that there are significantly fewer minima, and perhaps only one, has been a common approach in global optimization. \cite{Stillinger88, Stillinger90, Head-Gordon91, Piela94, Pillardy95}
The idea behind this approach is that global optimization is likely to be much easier on the smoothed landscape, and trivially so if there is only one minimum left.
Of this set of global optimization algorithms, particularly relevant to us are those that achieve this smoothing by changing a potential parameter. \cite{Stillinger90,Head-Gordon91,Pillardy95}

\begin{figure}
\centerline{\includegraphics[width=8.6cm]{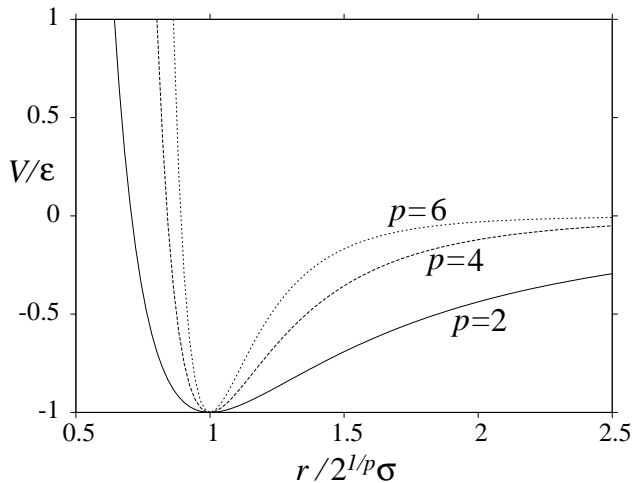}}
\caption{
The generalized Lennard-Jones potential (Eq.~\ref{eqtn:glj}) for $p=2,4$ and 6. .
}
\label{fig:vr}
\end{figure}

Here, we examine the evolution of the landscape of the 13-atom cluster for a generalized Lennard-Jones potential with parameter $p$, defined by
\begin{equation}
\label{eqtn:glj}
V = 4 \epsilon \left[ \left( \frac {\sigma} {r} \right)^{2p} - \left( \frac {\sigma} {r} \right)^{p} \right],
\end{equation}
\noindent where $r$ is the distance between two atoms, $\epsilon$ is the well-depth and $2^{1/p} \sigma$ is the equilibrium pair separation.
This is summed over all pairs of atoms to give the total potential energy of a configuration.
The range of the potential is determined by the parameter $p$, where for $p=6$, the potential is equivalent to the Lennard-Jones potential.
As $p$ decreases, the repulsion becomes less steep and the attraction has a longer range.
The net effect is that the width of the well increases, as shown in Fig.~\ref{fig:vr}.

The effects of $p$ (or the analogous parameter for the Morse potential) on the landscape have been well studied. \cite{Stillinger90, Hoare76, Hoare83, Braier90, Doye95b,Doye96,Doye96c,Doye97,Miller99,Miller99b}
At sufficiently small $p$, there is a single minimum on the landscape.
As $p$ increases, the number of minima increases rapidly, because potential wells surrounding the minima become narrower, allowing new minima to appear at higher energy.
Equivalently, as $p$ decreases, the lower energy wells become wider, thus swallowing up the higher energy minima.
Downhill barriers also increase as $p$ increases.
The net effect is that a smooth landscape at low $p$ becomes rougher.

In Section \ref{sec:cat}, we provide details of how the appearance and disappearance of minima and transition states can be described by catastrophe theory, \cite{Wales01,catbook} and the effect on the inherent structure network.
In Section \ref{sec:pa}, we describe the method used to detect preferential attachment.
In Section \ref{sec:results}, we discuss how the landscape for the 13-atom Lennard-Jones cluster evolves, and whether preferential attachment provides a useful description of that evolution.

\section{Catastrophes}
\label{sec:cat}

\begin{figure}
\centerline{\includegraphics[width=8.6cm]{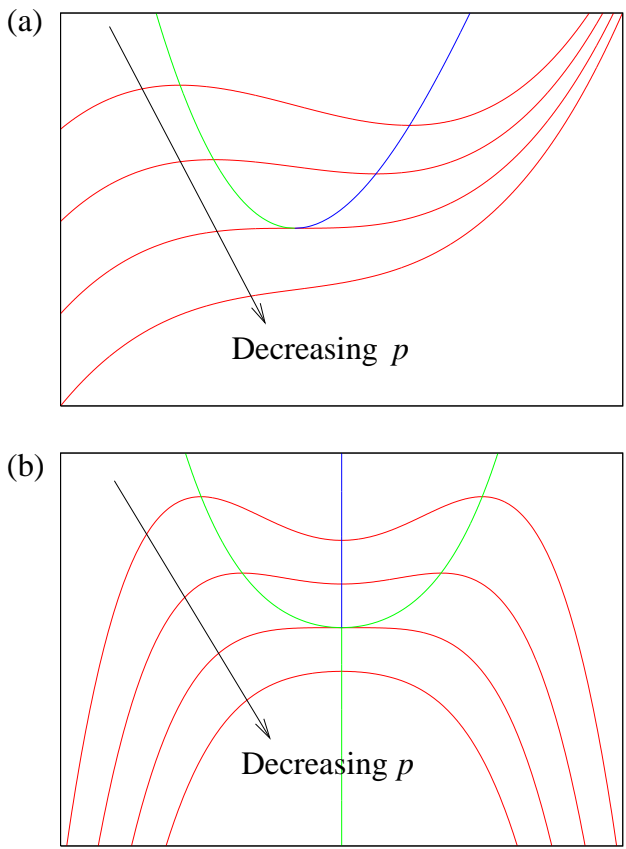}}
\caption{(Colour online)
Changes in a landscape close to a catastrophe.
The loci of the stationary points are also plotted.
(a) The fold catastrophe. 
Initially, there are two stationary points, a maximum and a minimum.
Upon varying a parameter, decreasing $p$ (increasing the range) in the case of the PEL, these stationary points move closer together until they collide and disappear.
(b) The cusp catastrophe. 
Initially, there is one minimum and two maxima, which collide upon increasing the range to leave a single maximum.}
\label{fig:cats}
\end{figure}

The two most common types of catastrophe for PELs are the fold and the cusp, illustrated in Fig.~\ref{fig:cats}.
In the fold, a minimum and a maximum move closer together as some parameter is varied, in this case as the range is increased.
At the catastrophe the two stationary points meet and disappear, leaving just a shoulder in the landscape.
The cusp in Fig.~\ref{fig:cats}(b) involves two symmetry-related maxima and one minimum.
Upon increasing the range, they move closer together until they collide, leaving one maximum.

Fig.~\ref{fig:cats} is a one-dimensional picture, whereas the PEL is high-dimensional.
However, these catastrophes can provide a good description of the evolution of one-dimensional `reaction coordinates' that link stationary points on the PEL. \cite{Wales01}
These reaction coordinates are defined in terms of steepest-descent pathways, i.e.~the paths follow the gradient downhill at every point, and thus there is no component of the gradient in the space orthogonal to the coordinate.
Hence the points where there is a maximum or minimum in the reaction coordinate correspond to stationary points of the PEL. 

We are interested in catastrophes involving minima and transition states, where minima are stationary points that are minimal in all directions, and transition states are maximal in one direction and minimal in all others, i.e.~the Hessian has one negative eigenvalue.
This leaves two possible scenarios.
In the first, the minima in Fig.~\ref{fig:cats} represent minima and the maxima transition states.
In the second, the minima in Fig.~\ref{fig:cats} represent transition states and the maxima index 2 saddles, which are maximal in one more dimension than transition states.
The cusp catastrophe can only occur when there is symmetry present, as the pathways either side of the central stationary point are equivalent.
Therefore, any stationary points connected by the pathways must be permutational isomers, where the system has the same structure but some atoms swap positions.
The low $p$ maximum must therefore connect two permutational isomers of the same minimum.

There are some subtleties in the way the inherent structure network has been defined.
Edges correspond to transition states, but not all transition states give rise to edges.
More than one transition state may connect the same pair of nodes, forming a multiple edge (ME), and some connect permutational isomers of the same minimum, forming a self-connection (SC).
ME/SCs are not considered in the inherent structure network as we are interested only in whether two minima are connected, not how many times, and we do not distinguish between permutational isomers of a minimum.
This choice is important for the network topology. \cite{Maslov02,Doye05b}

\begin{figure}
\centerline{\includegraphics[width=8.6cm]{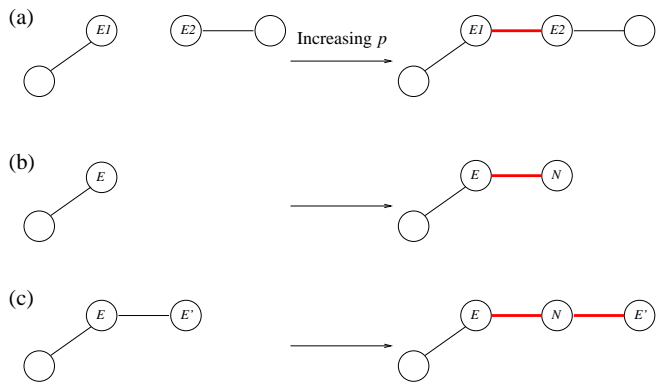}}
\caption{(Colour online)
Changes the inherent structure network that can occur on increasing $p$.
(a) Addition of an internal edge (shown in bold) connecting two existing nodes, $E1$ and $E2$.
(b) Addition of an external edge connecting a new node $N$ to an existing node $E$.
(c) Addition of an external edge via a cusp.
The initial transition state connects two permutational isomers, $E$ and $E'$, forming a self-connection.
The new node is inserted into this edge, forming connections to both permutational isomers.
However, because we ignore ME/SCs, the overall effect on the network topology is equivalent to that in (b).
}
\label{fig:net}
\end{figure}

The two catastrophes shown in Fig.~\ref{fig:cats} can affect the inherent structure network as shown in Fig.~\ref{fig:net}.
We classify catastrophes as the addition of either an internal or external edge to the network. \cite{Albert00b,Dorogovtsev01,Mattick05,Gagen05}
An internal edge involves the addition of an edge between two existing nodes.
This can occur on increasing $p$ via a fold catastrophe, creating a transition state and an index two saddle from an inflection point, or a cusp catastrophe, creating one transition state and two index two saddles from an index two saddle.
Both these processes occur along the boundaries between two basins of attraction.
The new transition state may connect two permutational isomers of the same minimum, in which case it would form an SC, or it may connect two minima that are already connected by a transition state, forming a ME, and therefore have no effect on the network topology.

An external edge involves the addition of a new node and one new edge.
This can either be from a fold catastrophe, an inflection point becoming a minimum and a transition state, or a cusp catastrophe, a transition state becoming a minimum and two transition states.
Symmetry restrictions on the cusp catastrophe affect the network as shown in Fig.~\ref{fig:net}(c).
The initial transition state connects two permutational isomers, therefore it forms an SC in the network and is not included in our representation.
A cusp inserts a minimum into this edge, that is connected to both permutational isomers.
As we only count one permutational isomer in the inherent structure network, the effect on the network topology is that one external edge is added.

\begin{figure}
\centerline{\includegraphics[width=8.6cm]{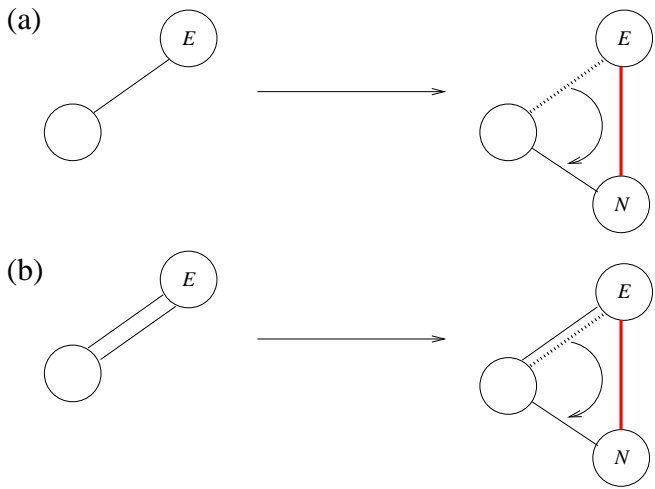}}
\caption{(Colour online)
The two major ways that the addition of an external edge (shown in bold) connecting new node $N$ to existing node $E$ can lead to an edge being rewired, thus increasing the initial degree of $N$.
In (a), the edge is rewired from $E$, thus reducing its degree.
In (b), the rewired edge was originally an ME, thus the degree of $E$ is not affected.
}
\label{fig:rew}
\end{figure}

When a new minimum appears on the landscape, an external edge is formed to one existing node by the new transition state that was involved in the catastrophe.
However, the change to the network topology can be more complicated.
The basin of attraction associated with a new minimum generated by a fold catastrophe has a finite area, and as such can cause reaction pathways from existing transition states to be rewired to connect to it if the existing reaction pathway passed through the region of configuration space now assigned to the basin of the new minimum (Fig.~\ref{fig:rew}).
The new node can therefore have initial degree greater than one, although there is only one new transition state.
These rewired transition states may be rewired from MEs and as such appear to be new edges in terms of the network topology.
Because such new edges connect to the new node, they are classified as external edges.

Although these are the most likely types of catastrophe, there may be other, more complicated catastrophes.
However, such catastrophes involve higher symmetries and are therefore rare.
Furthermore, we do not search specifically for folds and cusps, simply for stationary points on the PEL.
Starting from $p=6$, for which it is likely that all of the stationary points are known, we decrease $p$ (increasing the range of the potential) and record when stationary points disappear from the surface, and when transition states rewire to connect to different minima.
This information is then inverted to give the details of the network's growth.
While it is possible for a minimum or transition state to appear as the range is increased, this type of event is difficult to detect and is expected to be relatively rare, so is not included.
Rewiring of transition states is only considered when a minimum is involved in a catastrophe, as this is the most likely occasion for it to occur.

\section{Measuring Preferential Attachment}
\label{sec:pa}

In preferential attachment models, when a new node appears in the network, it is assumed to connect to an existing node with probability $\Pi$, depending on its degree $k$ as $\Pi \propto k^{\alpha}$.
The exponent $\alpha$ describes the strength of the preferential attachment.
In the original Barab\"asi-Albert model, \cite{Barabasi99} $\alpha$ must be one to give a scale-free network. \cite{Krapivsky00}
In more complex models, allowing for example new edges between a pair of existing nodes, \cite{Newman01b,Barabasi02,Jeong03,Eisenberg03,Roth05,Peltomaki05} other values of $\alpha$ can give rise to scale-free networks.

The exponent $\alpha$ can potentially be determined from time-resolved data for the growth of the network.
However, extracting the exponent is not straightforward for two reasons.
Firstly, network growth is typically stochastic.
A given preferential attachment rule can result in many different networks, and as such the exponent cannot be determined exactly.
Secondly, at each time step the probability of connecting to a node with degree $k$ is $\Pi = k^{\alpha}/\sum_ik_i^{\alpha}$.
The normalising sum in the denominator is time-dependent, and depends on the unknown exponent.
Approximations have been used previously, such as comparing the network at two different stages of its growth and assuming the normalising sum remains constant over this time period, \cite{Jeong03,Barabasi02,Roth05} or assuming that the normalisation sum is proportional to the size of the network. \cite{Newman01b}
By contrast, we use an iterative, self-consistent method that we recently introduced. \cite{Massen05c}
In this method, we estimate $\alpha$ in order to evaluate the sum, then use the results to obtain a better estimate for $\alpha$.

The number of edges gained by all nodes with degree $k$ at time $t$, $\Delta k$, depends on the number of nodes with degree $k$ at time $t$, $n_k(k,t)$, and is given by
\begin{equation}
\Delta k(k,t) = n_k(k,t)\Pi(k,t) = \frac{ n_k(k,t) k^{\alpha} } { \sum_ik_i(t)^{\alpha} } = \frac{ n_k(k,t) f(k) } {c_t(t)},
\end{equation}
\noindent where we define $f(k)=k^{\alpha}$ and $c_t(t)=\sum_ik_i(t)^{\alpha}$.
$f(k)$ is obtained by summing both sides over all time steps, thus avoiding problems from steps where $n_k(k,t)=0$, giving
\begin{equation}
\label{eqtn:fk}
f(k)= \frac { \sum_t {\Delta k(k,t)} } { \sum_t {n_k(k,t)/c_t(t)} }.
\end{equation}
\noindent $\alpha$ is then obtained from a linear least squares fit to $\log f(k) = \alpha \log k$, weighted by the number of terms in the sum in the denominator of Eq.~\ref{eqtn:fk}.
Logarithmic binning over $k$ reduces errors from degrees where $\sum_t {\Delta k(k,t)}=0$, which cannot be included in a log-log plot.
It has been shown that the resulting exponent is independent of the initial estimate for $\alpha$, and the method generally converges. \cite{Massen05c}
In tests on model networks grown with a known preferential attachment rule, the exponent obtained had an absolute error less than approximately 0.13 for $0.5 \le \alpha \le 1.5$, with a much lower error for $\alpha$ closer to 1.

We investigate the addition of internal and external edges independently.
For internal edges, the probability of a pair of nodes gaining an edge is assumed to depend on the product of their degrees as $\Pi(k_ik_j) \propto (k_ik_j)^{\alpha}$.
If multiple edges are not allowed in a network, when determining $\alpha$ only those pairs of nodes that are not connected should be used in the normalization sum $c_t(t)$.
This modification has been seen to greatly improve the exponent obtained, reducing the error from approximately 20\% to 1\% in the case that $\alpha=1$. \cite{Massen05c}
This approach can straightforwardly be applied to the evolution of the inherent structure networks, the only unusual feature being that the networks grow as a function of the potential parameter $p$, not time.

\section{Results}
\label{sec:results}

\subsection{Network properties}

\begin{figure}
\centerline{\includegraphics[width=8.6cm]{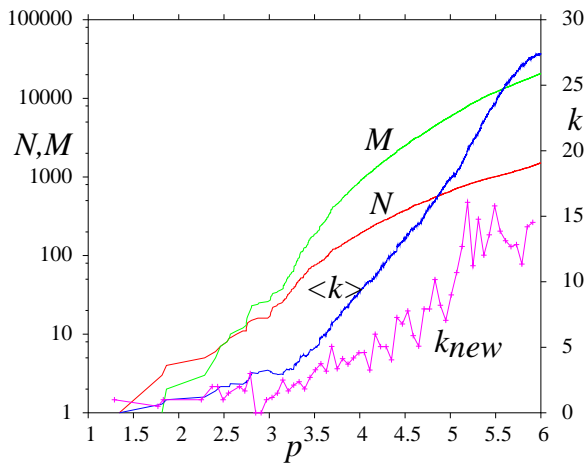}}
\caption{(Colour online)
Growth with $p$ of
the number of nodes, $N$, 
the number of edges, $M$,
the average degree, $\langle k \rangle = 2M/N$, and
the initial degree of a new node, $k_{new}$ (binned over $p$).
}
\label{fig:netp}
\end{figure}

The number of nodes and edges increase as the range parameter is increased towards $p=6$, at which the network contains 1509 nodes and 20\,691 edges.
As shown in Fig.~\ref{fig:netp}, the number of nodes increases roughly exponentially with $p$, i.e.~the network grows faster as $p$ approaches 6.
The number of edges also increases roughly exponentially with $p$.
As cusp catastrophes occur at either transition states or index 2 saddles, and fold catastrophes occur at inflection points, both of these will be more likely when the landscape is rougher (i.e.~the network is bigger).
In particular if $dN/dp \propto N$ exponential growth results.

The number of edges increases faster than the number of nodes, so the average degree increases, i.e.~the network is accelerating and becoming better connected.
This is due to two factors.
Firstly, the addition of internal edges.
Secondly, the initial degree of a new node increases with $p$.
If a node is added at higher $p$, it is likely to link to more existing nodes because there are more nodes in the network, and more edges to rewire.
Some of these edges are rewired from MEs, thus increasing the average degree.

\begin{figure}
\centerline{\includegraphics[width=8.6cm]{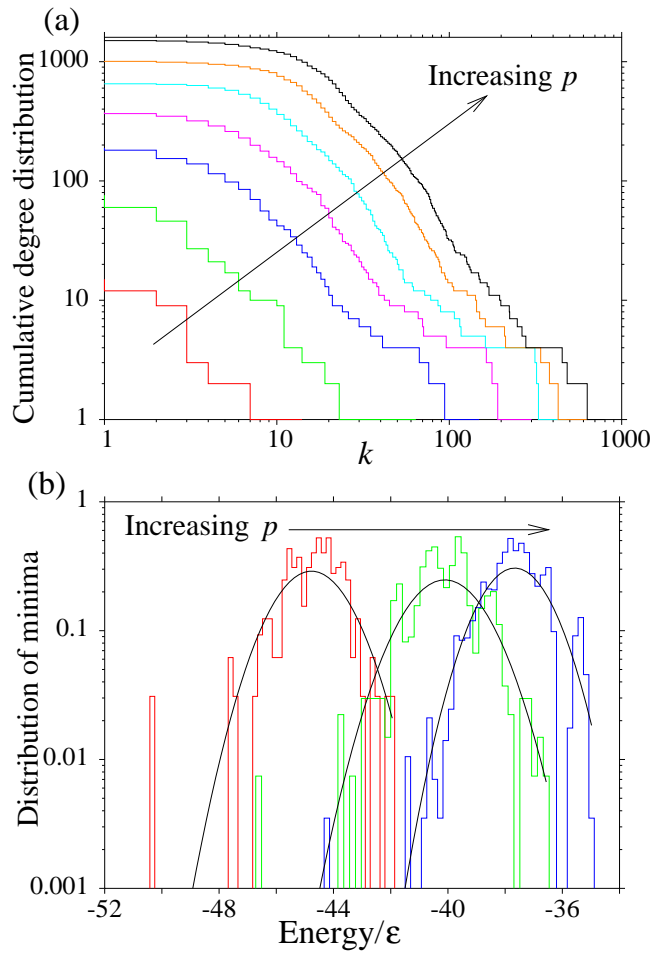}}
\caption{(Colour online)
(a) Cumulative degree distribution (i.e.~the number of nodes with degree greater than $k$) and 
(b) distribution of the energies of the minima 
at various values of $p$ during the growth of the network.
In (b), Gaussian best fits are also shown.
In calculating these best fits, the global minimum is excluded as it has especially low energy, due to the particularly stable icosahedral structure of the cluster. 
}
\label{fig:pk}
\end{figure}

\begin{figure}
\centerline{\includegraphics[width=8.6cm]{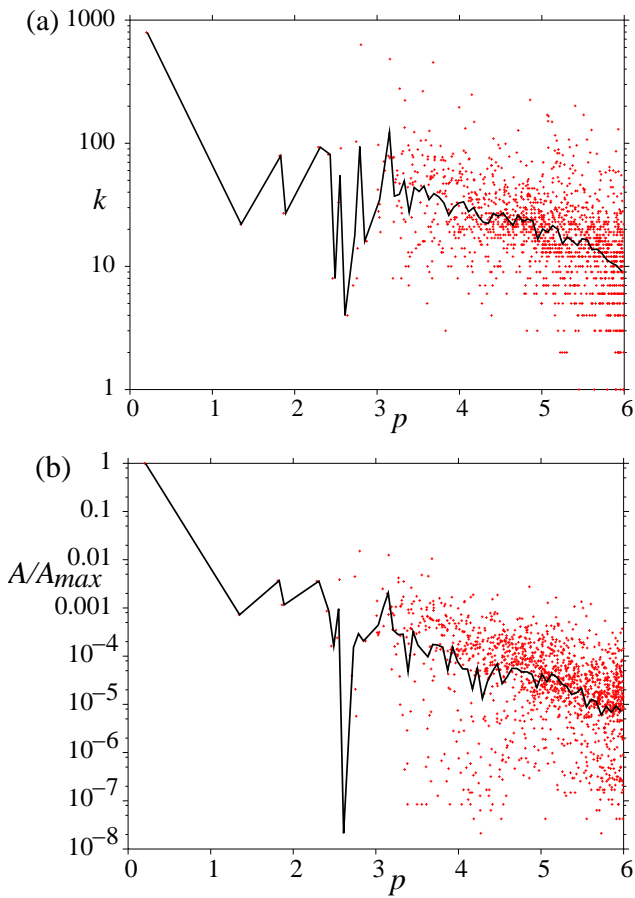}}
\caption{(Colour online)
Properties of a node at $p=6$ as a function of the value of $p$ when it was created.
(a) Node degree, $k$, and
(b) basin area, $A$.
Individual points correspond to individual nodes and 
the solid line is a binned geometric average.
Note that due to the exponential growth of the network, only 17 nodes, 1.1\% of the total, appear at $p \le 3$, where the correlation of the properties with 
$p$ is weaker. 
}
\label{fig:age}
\end{figure}

The degree distribution has a similar form throughout the growth process, as shown in Fig.~\ref{fig:pk}(a), indicating that the growth is fairly homogeneous in terms of network topology, and the network reaches a kind of steady state.
Furthermore, the distribution of energies of the minima is approximately Gaussian throughout the growth process (Fig.~\ref{fig:pk}(b)).
Although the initial degree of `newer' nodes (those added at higher $p$) is higher, the final degree at $p=6$ is lower than for the older nodes (Fig.~\ref{fig:age}(a)).
This ``age'' dependence of the degree is indicative of preferential attachment and has been seen previously, for example in a protein interaction network. \cite{Eisenberg03}
It should be noted that the network is only large enough to study degree distributions for $p \ge 3$.
In this region, the growth of the number of stationary points, and particularly the average degree, is also fairly steady (Fig.~\ref{fig:netp}).

\subsection{Preferential attachment}

\begin{table}
\caption{\label{table:exponents}
Exponents found for preferential attachment of external edges ($\Pi(k) \propto k^{\alpha}$) and internal edges ($\Pi(k_ik_j) \propto (k_ik_j)^{\alpha}$).
The number of data points used to obtain the exponents are also shown.
External edges are further broken down into those that involve addition of a new transition state and those that involve rewiring a transition state.
Rewired transition states are only considered if they involve an increase in degree of the existing node, i.e.~before the catastrophe, the transition state formed an ME.
}
\begin{ruledtabular}
\begin{tabular}{ccc}
Type of time step        & Exponent $\alpha$ & Number \\
\hline
External edges (new TS)     & 1.20     &  1\,199 \\
External edges (rewired TS) & 1.24     &  7\,275 \\
Internal edges              & 0.99     & 12\,274 \\
\end{tabular}
\end{ruledtabular}
\end{table}

The exponents that we obtained by applying the algorithm in Section \ref{sec:pa} are shown in Table \ref{table:exponents}.
They are close to one, indicating that preferential attachment is present in the growth of the network.
New edges are more likely to link to existing nodes with high degree, leading to a scale-free network.
The exponent is slightly greater than one for external edges both for those associated with new transition states, and those associated with rewired MEs.
When a new transition state appears on the PEL that gives rise to an external edge, the degree of the existing node to which it connects increases by one.
Simultaneously, some edges are rewired from existing nodes, increasing the initial degree of the new node.
However, due to the spatial nature of the PEL, these rewired edges are likely to be rewired from the existing node that is connected to the new transition state (see Fig.~\ref{fig:rew}).
In fact, on average, when such an external edge is added, the existing node at the end of the edge loses six edges to rewiring, whilst only gaining one.
This opposes preferential attachment by taking edges from high-degree nodes.
However, this effect is cancelled out by the preferential attachment associated with external edges that simultaneously appear due to the rewiring of MEs.
Furthermore, the frequency with which external edges are added is lower than that for internal edges (Table \ref{table:exponents}) and the appearance of new minima increases the available basin boundary along which new internal edges can be generated.

\begin{figure}
\centerline{\includegraphics[width=8.6cm]{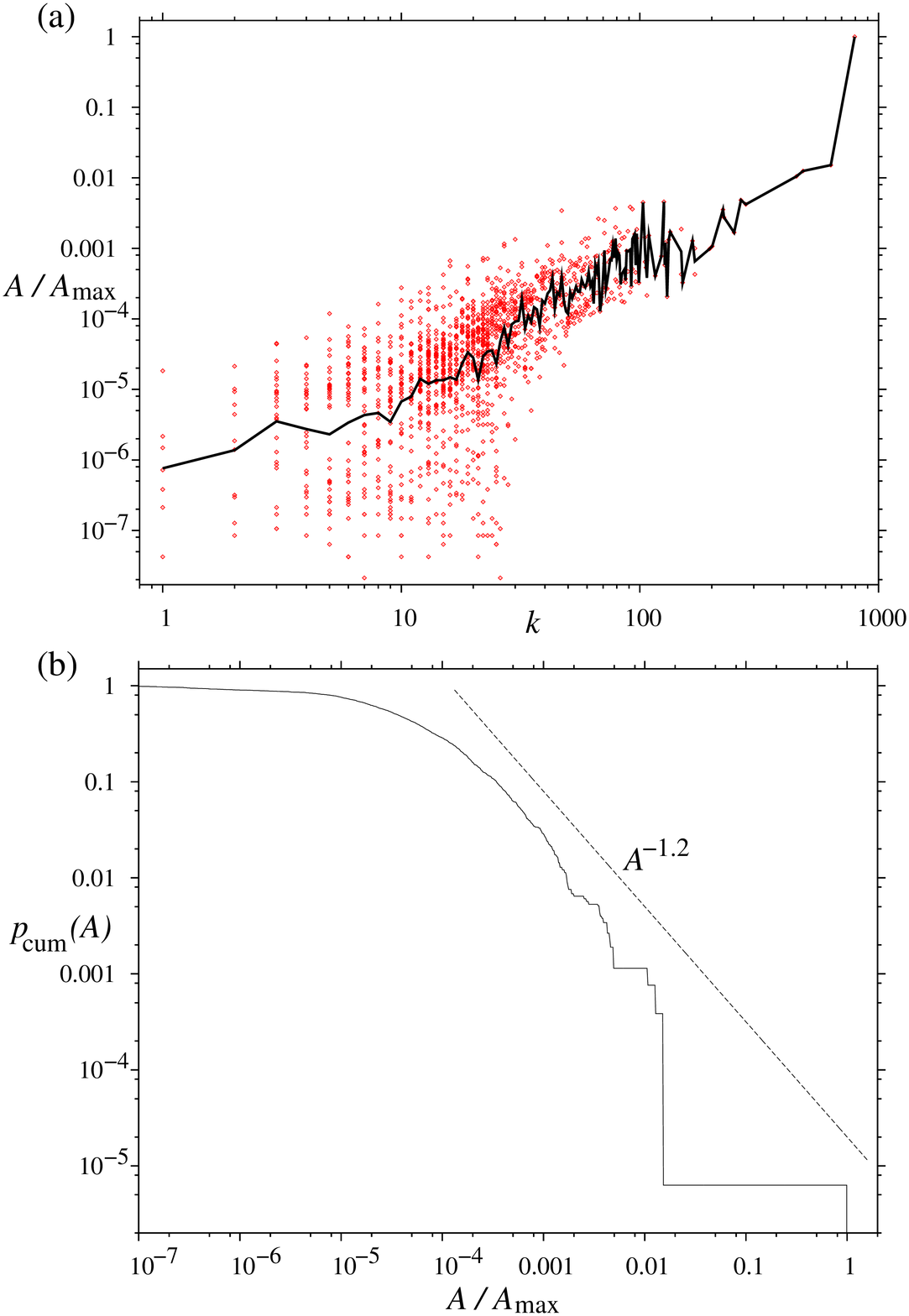}}
\caption{(Colour online)
(a) Degree dependence of the area of the basin of attraction of a minimum for
LJ$_{13}$. 
Individual points correspond to individual minima and
the solid line is a geometric average.
(b) The cumulative basin area distribution. In (b) a power-law form
corresponding to $p(A)\sim A^{-2.2}$ (the exponent of the cumulative 
distribution is greater by one) has also been plotted.
}
\label{fig:area}
\end{figure}

Internal edges do increase the degree of high-degree nodes, as expected for preferential attachment, and the exponent for this process is very close to one.
Since these ``time steps'' dominate the network growth, the preferential attachment associated with the addition of internal edges seems to be the main driving force for the scale-free topology of the final network.
As internal edges are generated by catastrophes occurring along the basin boundaries, the larger, older minima are likely to  garner more internal edges, hence the preferential attachment.
Indeed, a correlation between degree and basin area is shown in Fig.~\ref{fig:area}(a).

To probe the connection to the basin area distribution in more detail, we have measured the areas of the basins of attraction directly.
This task is achieved simply by choosing random points in configuration space and then performing a local minimization.
The area of a basin is then proportional to the number of points that lead to the corresponding minimum after minimization.
Newer minima have smaller basins of attraction in the final PEL at $p=6$ (Fig.~\ref{fig:age}(b)).
When new minima appear on the landscape they are likely to do so near to the edge of a basin of attraction.
Near to the centre of the basin, the basin behaves harmonically, as quadratic terms dominate the Taylor expansion of the energy about the minimum.
Catastrophes are driven by higher-order anharmonic terms, and so will only occur at sufficient distance from the minimum.

At the exact moment at which the catastrophe occurs and the minimum appears in the network, the minimum and the new transition state also involved in the catastrophe are at the same point.
For the fold catastrophe, the basin associated with this minimum initially has zero extent in the direction of the transition state, but both in the opposite direction and all those directions perpendicular to this reaction coordinate, it is of finite extent.
As $p$ increases, the distance between the transition state and minimum will increase, albeit relatively slowly --- close to the catastrophe this distance can be approximated by $\sqrt{6\Delta V/\lambda}$, where $\Delta V$ is the barrier, and $\lambda$ is the curvature at the minimum and transition state along the reaction path. \cite{Wales01}
Therefore, the area of the new basin will probably initially increase, but it is always likely to be smaller than the basins of the older minima between which it is sandwiched.
Hence, as $p$ increases and there are more minima on the landscape, the new minima will have increasingly small basins, as they are more constrained by the more closely spaced basins.

Combining this effect with the growth of the network leads to a power-law area distribution.
New basins that appear at higher $p$ have smaller basin areas.
Due to the exponential increase of $N$ with $p$, there are many of these new, small basins.
Furthermore, the nature of the catastrophes means that the new basins are likely to appear at or near the boundaries of larger, existing basins.
For this reason, the distribution of minima in configuration space will be inhomogeneous, with minima concentrated around the edges of the larger basins in a manner that is somewhat similar to the spatial distribution of hyperspheres in the Apollonian packing (Fig.\ \ref{fig:apollo}.

Both the number of minima at a given $p$ and the area of a basin (at $p=6$) that appeared at $p$ show an approximately exponential dependence on $p$, i.e. $N(p) \sim \exp(\nu p)$ and $A(p) \sim \exp(-\alpha p)$.
Hence we can derive an expression for the area distribution.
The number of basins with area $A$ at $p=6$ is given by
\begin{eqnarray}
n(A)dA & =    & \frac{dN}{dp} dp \nonumber \\ 
n(A)   & \sim & \nu \exp(\nu p) \left| \frac{dp}{dA} \right|  \sim  \frac{\nu \exp(\nu p)}{\alpha \exp(-\alpha p)}.
\end{eqnarray}
Substituting for $p$ via $A(p) \sim \exp(-\alpha p)$, we arrive at
\begin{eqnarray}
n(A)   & \sim & \frac{\nu}{\alpha} A^{-1-\nu/\alpha}. 
\end{eqnarray}
\noindent This mechanism for generating a power law has been termed combination of exponentials by Newman. \cite{Newman05}
The observed exponents are $\nu \approx 1.1$, $\alpha \approx 1.2$, giving $P(A) \sim A^{-1.9}$.
This result is in line with the observed approximate power-law tail to the 
basin area distribution for LJ$_{13}$ (Fig.\ \ref{fig:area}(b)).
The exponent is also close to that for the area distributions observed for liquids and for Apollonian packings. \cite{Massen05}
For the latter the exponent is exactly $-2$ in the limit of high dimensionality.
Interestingly, the analogies to the Apollonian packing go farther, as the power-law in the Apollonian case can also be considered to arise from a combination 
of exponentials.
Namely, for the two-dimensional Apolloinian packing, it has been shown that 
both the number of disks and their area depend exponentially on the number of 
iteration steps used to generate the packing, 
and that, furthermore, new disks preferentially attach to those
with high degree.\cite{Doye05}

\section{Conclusion}
\label{conc}

We have used techniques from network theory to study the evolution of a PEL on going from a simple form with a single minimum to a complex landscape with a large number of minima as the range of the underlying interatomic potential is varied.
In particular, we have examined how the inherent structure network, which provides a picture of the connectivity of the PEL, changes.
New edges in this network are seen to preferentially attach to minima with a larger number of connections, helping to explain the scale-free nature of the inherent structure network.
This preferential attachment is very close to linear, and can be understood by considering the underlying catastrophes that lead to the appearance of new minima and transition states on the PEL.
Each catastrophe that occurs, as well as increasing the degree of a minimum, also makes the surface rougher, leading to further catastrophes.
These catastrophes are also more likely to occur at or near boundaries, therefore they are more likely to occur near large basins, which have high degree.

Consideration of this growth process leads to further insight into the nature of the division of the PEL into basins of attraction.
Previous work has shown that, similar to the Apollonian packing, PELs have power-law distributions for the hyperareas of these basins of attraction, suggesting that the basins tile the landscape in a fractal-like manner.
The current results are able to shed light on this puzzling character of the PEL.
During the evolution of the PEL small new basins appear near the boundaries of larger existing boundaries, as in the generation of the Apollonian packing.
The network grows exponentially, because catastrophes make the network larger and the surface rougher, enabling further catastrophes.
The combination of this exponential growth and the smaller size of newer basins leads to a power-law area distribution, in agreement with observed results for PELs.

\end{document}